\documentclass[twocolumn,showpacs,aps]{revtex4}

\usepackage{graphicx}
\usepackage{amssymb}
\usepackage{array}
\usepackage{multirow}

\begin{document}

\title{Hypersound damping in vitreous silica measured by picosecond acoustics}

\author{A. Devos$^{1}$, M. Foret$^{2}$, S. Ayrinhac$^{2}$, P. Emery$^{3}$,  and B. Ruffl\'e$^{2}$}

\affiliation{$^{1}$Institut d'Electronique, de Micro\'electronique et de Nanotechnologie, UMR CNRS 8250,\\
BP 69, Avenue Poincar\'e, F-59652 Villeneuve d'Ascq Cedex, France\\
$^{2}$Laboratoire des Collo\"{\i}des, Verres et Nanomat\'eriaux, UMR CNRS 5587,\\
Universit\'e Montpellier 2, F-34095 Montpellier Cedex 5, France\\
$^{3}$ST Microelectronics, 850 rue Jean Monnet, F-38926 Crolles Cedex, France}

\date{October 2, 2007}

\begin{abstract}
The attenuation of longitudinal acoustic phonons up to frequencies nearing 250 GHz is measured in vitreous silica with a picosecond optical technique.
Taking advantage of interferences on the probe beam, difficulties encountered in early pioneering experiments are alleviated.
Sound damping at 250~GHz and room temperature is consistent with relaxation dominated by anharmonic interactions with the thermal 
bath, extending optical Brillouin scattering data.
Our result is at variance with claims of a recent deep-UV experiment which reported a rapid damping increase beyond 100 GHz.
A comprehensive picture of the frequency dependence of sound attenuation in $v$-SiO$_2$ can be proposed.
\end{abstract}

\pacs{61.43.Fs, 63.50.+x, 78.47.+p} 
\maketitle

Despite considerable recent activity, high frequency sound attenuation in glasses remains a controversial issue. 
Several processes lead to sound-wave damping in glasses.
At temperatures $T$ above the quantum regime, the most important ones are: thermally activated relaxations \cite{And55,Jae76}, interaction with thermal vibrations \cite{Fab99}, scattering from disorder \cite{Kle51,Ell92,Sch93}, 
and resonance with low-lying modes \cite{Kar85,Buc92}.
Sorting these out in function of $T$ and the sound-wave angular frequency $\Omega $ proves to be difficult.

In acoustic experiments, one measures the energy attenuation coefficient $\alpha$, or the linewidth of Brillouin scattering lines $\Gamma=\alpha v$, where $v$ is the sound velocity and $\Gamma$ is in angular frequency. 
At low $\Omega $ and sufficiently low $T$, it is generally agreed that $\alpha $ is dominated by thermally activated relaxation (TAR) of structural ``defects''.
In vitreous silica, $v$-SiO$_2$, a strong broad attenuation peak is observed around 50 K in ultrasonic experiments \cite{And55}.
This is well described by the phenomenology of TAR \cite{Jae76}, although the exact nature of the ``defects'' is not clear. 
The damping depends on their strength, density, and distribution. 
A second damping mechanism, which dominates in crystals, is anharmonicity.
In the Akhiezer description, sound modulates the population of thermal bath modes which relax via anharmonic interactions \cite{Akh39}. 
This relaxation can be characterized by a mean thermal life time $\tau_{\rm th}$. 
As $\tau_{\rm th}(T)$ is generally short, anharmonicity is expected to dominate at higher frequencies. 
It leads then to a Brillouin linewidth $\Gamma_{\rm anh} \propto \Omega^2$ as long as $\Omega \tau_{\rm th} << 1$. 
This damping contribution increases faster with $\Omega $ than the $\Gamma_{\rm TAR}$ relaxation, as the latter should 
saturate \cite{Vac05}. 
For $v$-SiO$_2$, detailed estimates show that at a typical Brillouin light-scattering (BLS) frequency ($\approx 35$ GHz) and room temperature, TAR and anharmonicity mechanisms provide approximately {\small 1/3} and {\small 2/3} of the observed longitudinal acoustic (LA) attenuation, respectively \cite{Vac05}.

The existence of a third strong damping mechanism is evident from the universal plateau 
in the thermal conductivity of glasses around 5 K, which
implies that at high $\Omega$ the acoustic damping increases dramatically \cite{Zel71}. 
Based on the kinetic theory of thermal transport, $\alpha$ must increase at least as $\Omega^{4}$ to produce 
a plateau \cite{Ran88}.
Using inelastic x-ray scattering (IXS), direct spectroscopic evidence for such a dependence, 
$\Gamma \propto \Omega^4$, was recently obtained for the LA waves of two network glasses: 
densified silica \cite{Ruf03} and a lithium borate glass \cite{Ruf06}. 
However, IXS experiments being in practice limited to scattering vectors $Q > 1 $ nm$^{-1}$, the $(\Omega, Q)$ region 
where the Brillouin linewidth should exhibit this rapid increase remains inaccessible in most cases, 
{\em including } silica.
That this regime exists also in $v$-SiO$_2$ is fully endorsed by the observation that BLS linewidths do not extrapolate smoothly to IXS measurements, as shown below. 
The origin of this attenuation is currently debated. 
On the one hand, one might assume scattering of the sound waves by frozen-in disorder, {\em e.g.} by local 
elastic constant fluctuations \cite{Sch93,Sch06}. 
On the other hand, there can be a resonant interaction of sound waves with quasi-local vibrations \cite{Buc92,Par01}.
Recent quantitative estimates rather support the second mechanism \cite{Ruf08}.
In both cases, the explanation links to the boson peak, {\em i.e.} to an excess of low-frequency modes over the Debye expectation, another controversial feature of glasses. 
The above was recently challenged by deep-UV Brillouin scattering experiments claiming an unexpected sharp 
increase in the acoustic attenuation of silica at room-$T$ already around 100~GHz \cite{Mas06}, 
{\em i.e.} much below the boson peak frequency of $\sim 1$ THz.
These unsettled questions clearly call for more sound-absorption data between 100 GHz and 1 THz.

In this Rapid Communication, we report new measurements of the LA attenuation in $v$-SiO$_2$ thin films in this 
crucial frequency region using a sensitive modification of the {\em picosecond ultrasonic} technique 
(PU) pioneered in the 1980s \cite{Tho84}. 
Difficulties encountered in the early work \cite{Zhu91,Mor96} can be alleviated using the present data 
acquisition and analysis. 
Specifically, we find damping coefficients approximately one third those in \cite{Zhu91}, in line with available BLS data. 
The relaxation via anharmonic interactions with the thermal bath appears to be the dominant broadening mechanism at 300~K 
for $\Omega / 2 \pi$ between $\sim 190$ and $\sim  240$ GHz.

\begin{figure}
\includegraphics[width=6.5cm]{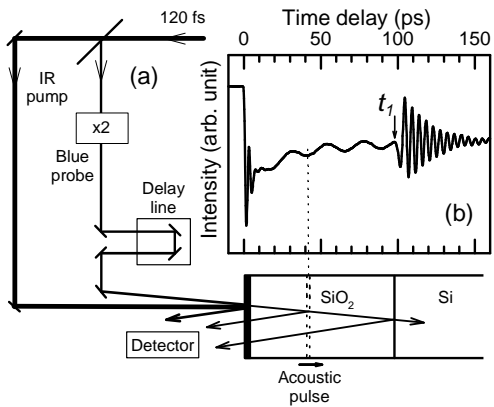} 
\caption{(a) Schematic diagram of the experiment; (b) Reflected intensity as a function of 
time delay obtained on an Al/SiO$_2$(600 nm)/Si sample using a 820 nm pump. \label{fig1}} 
\end{figure}

The samples are $v$-SiO$_2$ layers of four thicknesses from 300 to 1200 nm 
deposited on a (100) silicon substrate by low pressure chemical vapor deposition.
This produces layers with macroscopic properties very similar to those of bulk fused silica. 
The layers contain $\sim 3000$ ppm OH plus a small amount of hydrogen, below 500 ppm.
An Al film of thickness $d \simeq 12$~nm is evaporated on top of the samples in a single batch, to obtain identical films on all of them.
We use the standard pump-probe setup sketched in Fig.~1(a).
The light source is a titanium:sapphire oscillator tunable in the 700-980 nm wavelength range, producing 120 fs pulses. 
The pump is directly derived from the oscillator, whereas the probe is frequency doubled and time delayed. 
The pump and probe fall on the sample surface at near normal incidence, focused to a spot 
of $\sim 20$~$\mu$m diameter. 
The pump is absorbed by the thin Al film which expands launching a short longitudinal strain pulse into the sample.
The generated acoustic pulse is a plane wave packet of $\sim 20$ nm spatial extent
whose Fourier spectrum peaks around 250~GHz, following \cite{Tho86}.
This pulse undergoes a reflection at the layer/substrate interface producing an echo in the Al film which can be detected by the delayed probe.
In the early work by Zhu, Maris, and Tauc \cite{Zhu91}, the frequency dependence of the SiO$_2$ layer attenuation was determined from the decay of successive echoes.
A tungsten substrate was used to produce numerous and strong echoes.
Here, we use an interferometric method \cite{Tho86a,Lin91} to compare the strength of high frequency components of the acoustic pulse 
reaching the Si substrate in function of the layer thickness \cite{Dev04,Dev05,Eme06}.

\begin{figure} 
\includegraphics[width=8cm]{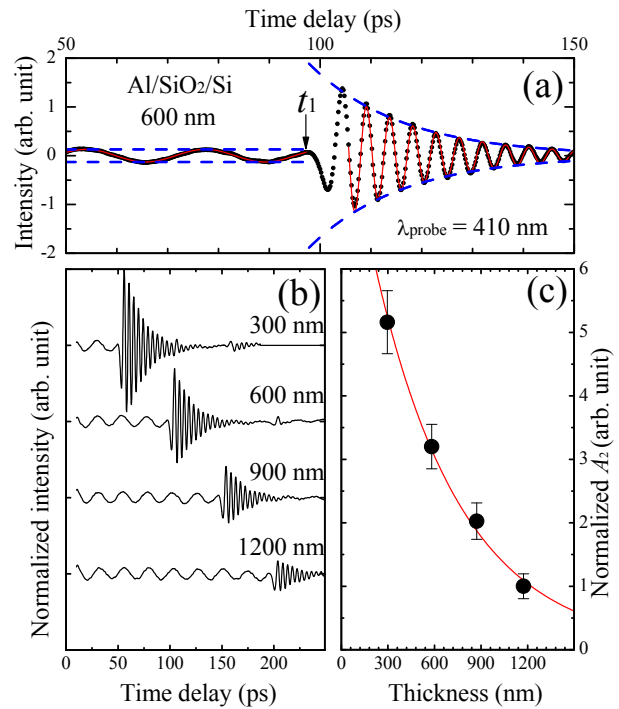} 
\caption{(Color online) (a) Typical fits (solid lines) of the signal intensity; 
(b) Intensities normalized to the $A_1$ amplitude, labeled by the nominal SiO$_2$-layer thickness;
(c) The normalized $A_2$-value as a function of the SiO$_2$-layer thickness. \label{fig2}} 
\end{figure}

The acoustic pulse is detected by means of the time-delayed probe as indicated in Fig.~1(a).
The main part of the probe is reflected at the fixed interfaces. 
A weaker component is reflected by the moving strain pulse, according
to the Bragg relation, $q = 2 k \cos\theta$ \cite{Tho86a},
where $q$ is the wave vector of the LA phonon, and $k$ and $\theta $ are the wave vector and
the angle of incidence of the probe light, respectively, both inside the material.
The two reflections interfere leading to oscillations
with the time delay $t$. The period is 
$$T = \lambda_{\rm probe}/(2 n v \cos\theta) \; \; ,$$
where $\lambda_{\rm probe}$ is the probe wavelength, $n$ is the refractive index at $\lambda_{\rm probe}$, and $v$ is the LA sound velocity.

A typical signal is shown in Fig.~1(b).
At zero delay, the intensity exhibits a sudden step followed by a long decay related to the thermal response of the Al film. 
Up to the transit time $t_1 = 98$~ps, a sinusoidal modulation of long period, $T_1 = 23.5 \pm 0.1$~ps is superposed to this background.
With $n = 1.470$ \cite{Mal65}, $T_1$ leads to the correct LA sound velocity of $v$-SiO$_2$, $v = 5940 \pm 25$~m/s. 
Beyond $t_1$, one observes strongly damped oscillations of smaller period, $T_2 = 4.57 \pm 0.02$~ps, in accordance with the refractive index 5.32 \cite{Asp83} and the LA sound velocity 8445~m/s \cite{McS53} of the Si(100) substrate. 
The frequency of the Fourier component probed in the SiO$_2$ layer is $1/T_1 = 42.5$~GHz while it is much higher, $1/T_2 = 219$~GHz, in the silicon substrate. 
One also observes during the first picoseconds a superimposed damped oscillation. 
This is the strain pulse partially bouncing back and forth within the Al film. 
From its transit time, $1.9 \pm 0.1$~ps, we calculate an Al-film thickness of $12.2 \pm 0.6$~nm, the same for all samples. 
This is an important point as the frequency distribution of the acoustic pulse is controlled by this thickness and our analysis assumes identical pulses in all samples.

For the analysis the background is first subtracted from the raw data. 
Typical signals after subtraction are shown in Fig.~2(a,b) \cite{fit}. 
The oscillations appear undamped in $v$-SiO$_2$ because both the light and sound absorptions are negligible at those frequencies.
In contrast, a strong damping rate of the oscillations, $5.51 \times 10^{-2}$~ps$^{-1}$, is observed in silicon, in agreement with the high light absorption coefficient of Si at $\lambda_{\rm probe}$ \cite{Asp83}.
We call $A_1$ and $A_2$ the initial oscillation amplitudes in the SiO$_2$ layer and in the Si substrate, respectively.
The acoustic attenuation of the layer at the high frequency, $1/T_2$, can be obtained measuring the decrease of $A_2$ with increasing layer thickness, as illustrated in Fig.~2(b)-(c).
For normalization, the signal divided by $A_1$ is plotted in Fig.~2(b).
To correctly normalize $A_2$ one must include in $A_1$ both the Stokes and anti-Stokes processes, plus the effect of the Fabry-Perot
interferometer formed by the $v$-SiO$_2$ layer which affects both $A_1$ and $A_2$.
Finally, the normalized $A_2$ value  follows an exponential decay with increasing SiO$_2$ thickness with an attenuation 
constant $ 1.80 \pm 0.15$~$\mu {\rm m}^{-1}$, from which we deduce the energy mean free path, $\ell \equiv \alpha^{-1}$, of LA phonons in the silica layer to be 280~nm at 219~GHz and room temperature.

\begin{figure}
\includegraphics[width=8cm]{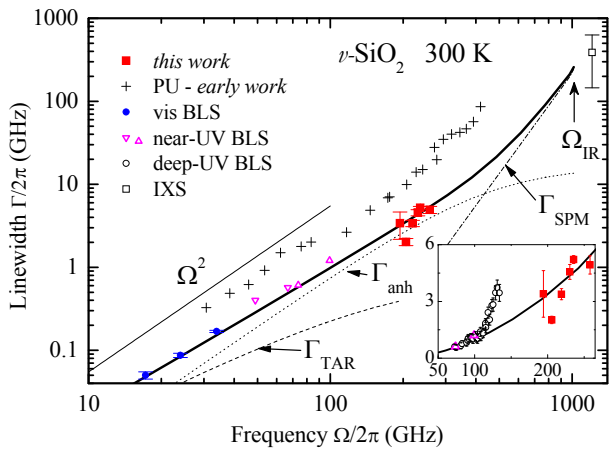}
\caption{ (Color online) Full widths $\Gamma(\Omega)$ of the Brillouin peaks deduced from different spectroscopies, as indicated. The full line is the estimated sum of TAR (dotted line), anharmonicity (dashed line) and SPM (dot-dashed line) contributions as explained in the text. The thin solid line of slope 2 is a guide for the eye. The inset illustrates the unexpected sudden rise of linewidth at $\sim$ 100 GHz reported in Ref. 18. \label{fig3}} 
\end{figure}

Tuning the laser wavelength we explored the LA phonon attenuation over the range 190-250 GHz. 
Our results are shown in Fig.~3 together with the early PU data \cite{Zhu91}, and with the available BLS \cite{Vac06, Mas04, Ben05} and IXS \cite{Mas97} data. 
To compare these results all are presented as Brillouin full widths. 
Also shown in Fig.~3 are the predictions of the most widely accepted models for the three main damping mechanisms. 
As explained above, the thermal activation of relaxing defects (TAR) accounts for the sound attenuation at sonic and ultrasonic frequencies while the {\em network viscosity} or anharmonicity takes over at the BLS frequencies.
The dotted and dashed lines in Fig.~3 illustrate the expected contribution of the TAR and the anharmonicity broadening mechanisms, respectively, following \cite{Vac05}, on the basis of experiments covering 7 decades in frequency \cite{TAR}.
Here, the curves are extrapolated up to the THz frequency range.
Moreover, one expects in this frequency range the onset of the sound attenuation $\propto \Omega^4$ discussed above.
This strong growth can lead rapidly to the end of acoustic plane waves at the Ioffe-Regel limit, which does provide an explanation 
for the universal low-$T$ plateau in the thermal conductivity \cite{Zel71}. 
In $v$-SiO$_2$ this limit is anticipated at $ \Omega_{IR} \sim 1$~THz, just at the boson peak maximum \cite{Ruf07b}. 
It corresponds to a wave vector $q \simeq 1$~nm$^{-1}$ for the LA mode, a region which is not 
attainable with current IXS spectrometers.
The full square in Fig.~3 obtained at 1.5~nm$^{-1}$ is, in fact, the lowest accessible IXS point 
measured at room-$T$ in $v$-SiO$_2$ \cite{Mas97}.
Within the framework of the soft potential model (SPM) \cite{Par94}, a quantitative prediction for the 
$\Gamma\propto\Omega^4$ behavior can be calculated \cite{Ram97}, shown as the dash-dotted line.
This $\Omega^4$ dependence is expected to terminate at the Ioffe-Regel limit, indicated as an arrow in Fig. 3.
The solid line is simply the sum of the three separate contributions.

In the low frequency part, it can be seen that the different $\Omega$ and $T$ dependences of $\Gamma_{\rm TAR}$ and $\Gamma_{\rm anh}$ lead to a non-trivial variation of $\Gamma$ with $\Omega$.
The departure from the  $\Omega^2$ behavior is indeed evident for the BLS results over the frequency range 20-100 GHz.
At higher frequencies, our PU data become dominated by the $\Omega ^2$ dependence of $\Gamma _{\rm anh}$, in remarkable 
agreement with the prediction.
The early PU data follow roughly the predicted dependence shown by the solid line, although it is obvious now that the absolute value was systematically overestimated.   
Several sources of error are in fact discussed in \cite{Zhu91}, the most serious one being possibly the corrections for losses at the interfaces.
The above scenario was challenged in a recent Letter reporting deep-UV Brillouin scattering data in which a dramatic increase of linewidth is  observed in the narrow frequency range 100-130 GHz as shown in the inset of Fig. 3 \cite{Mas06}.
The authors interpreted it as the onset of a new acoustic damping at a scattering vector $Q_{1} \simeq 0.11$~nm$^{-1}$ up to 
$Q_{2} \simeq 0.15$~nm$^{-1}$.
They associated it with Rayleigh scattering by disorder at the huge length scale, $2 \pi /Q_2 $.
This is surprising as it is generally thought that the scale of disorder in a glass does not exceed a few nanometers.
It is now clear from our results that no such damping onset occurs up to at least 250 GHz at room temperature in $v$-SiO$_2$.
The increase of the spectral linewidths observed in the deep-UV Brillouin scattering measurement could somehow be 
related with the optical absorption edge of $v$-SiO$_2$, in which case it is not an acoustic attenuation.

Above the Ioffe-Regel limit, there also exist IXS data, {\em e.g.} in \cite{Mas97}.
As shown in Fig. 3, anharmonicity damping saturates in that region so that it cannot explain the values observed in IXS giving an indirect evidence for the rapid growth of attenuation.
The onset of the $\Gamma\propto\Omega^4$ should occur around 400 GHz at room temperature in silica.
Above $\Omega_{IR}$, the exact nature of the excitations, which follow an {\em apparent} damping $\Gamma \propto \Omega ^2$, is so far unknown.
The fact that IXS spectra are observed and can be approximately adjusted to a damped harmonic oscillator lineshape is definitely not a proof that these excitations are propagating plane waves.
As discussed in \cite{Leo06}, elastic heterogeneities are seen in silica simulations at the boson peak scale.
A likely view is that beyond the Ioffe-Regel limit the acoustic modes strongly mix with quasi-local vibrations \cite{Ruf08} and thereby 
become diffusive, as supported by other simulations \cite{Tar99}.

In conclusion, we report new sound-attenuation coefficients in vitreous silica in the crucial 
frequency range between 100 GHz and 1 THz.
These new data points prolongate smoothly lower frequency Brillouin light scattering information.
The results fall on a line calculated independently on the basis of the three main damping contributions 
active in this frequency--temperature domain.
We observe that beyond 100~GHz up to at least 250~GHz it is anharmonicity that
dominates sound damping at 300~K in silica.
This implies that the regime of strong growth of the attenuation should arise beyond this frequency range, exactly on the low frequency side of the boson peak in silica.
Thanks to the present experimental breakthrough one can now envisage to observe the onset of the expected $\Omega ^4$
dependence.
As the latter mechanism should be $T$-independent, the onset is expected to move down to an accessible frequency range 
as $T$ is decreased reducing the anharmonicity damping. 
This will be a future crucial test of the overall picture.

The authors acknowledge invaluable discussions with R. Vacher and E. Courtens.


\begin{thebibliography}{99}
\bibitem{fit} We exclude from the fit the region of few ps width where the acoustic pulse passes through the SiO$_2$/Si interface.
\end{thebibliography}
\end{document}